\documentclass[aps,preprint]{revtex4}
\usepackage{amsmath}
\usepackage{epsfig}
\usepackage{float}
\usepackage{dcolumn} 

\begin{document}
\title{\bf \Large{A new solvable complex PT-symmetric potential}}  
\author{Zafar Ahmed$^1$, Dona Ghosh$^2$, Joseph Amal Nathan$^3$}
\affiliation{$~^1$Nuclear Physics Division, Bhabha Atomic Research Centre, Mumbai 400 085, India \\
$~^2$Astavinayak Society, Vashi, Navi-Mumbai, 400703, India \\
$~^3$Reactor Physics Design Division, Bhabha Atomic Research Centre, Mumbai 400085, India}
\email{1:zahmed@barc.gov.in, 2: rimidonaghosh@ gmail.com, 3:josephan@barc.gov.in}
\date{\today}
\begin{abstract}
\noindent
We propose a new solvable one-dimensional complex PT-symmetric potential as $V(x)= ig~ \mbox{sgn}(x)~ |1-\exp(2|x|/a)|$ and study the spectrum of $H=-d^2/dx^2+V(x)$. For smaller values of $a,g <1$, there is a finite number of real discrete eigenvalues. As $a$ and $g$ increase, there exist exceptional points (EPs), $g_n$ (for fixed values of $a$) causing a scarcity of real discrete eigenvalues, but there exists at least one. We also show these real discrete eigenvalues as poles of reflection coefficient. We find that the energy-eigenstates $\psi_n(x)$ satisfy (1): PT$\psi_n(x)=1 \psi_n(x)$ and (2): PT$\psi_{E_n}(x)=\psi_{E^*_n}(x)$, for real  and complex energy eigenvalues, respectively. 

\end{abstract}
\maketitle
PT-symmetric quantum mechanics [1,22,24] which started  theoretically [1-10,12] has penetrated well into the experimental and technological domains [11,14-16]. In this part of quantum mechanics, one considers  non-real, non-Hermitian  Hamiltonians which are invariant under the joint action of Parity ($P: x \rightarrow -x$) and Time-reversal ($T: i \rightarrow -i)$. Even the most simple Hamiltonian $H=-\frac{d^2}{dx^2}+V(x)$ corresponding to the Schr{\"o}dinger equation for these potentials has given 
astonishing results. Based on numerical computations, Bender and Boettcher [1] conjectured that the spectrum of $V_{BB}(x,\epsilon)=x^2(ix)^\epsilon$ was entirely real when $\epsilon \ge 0$. This conjecture has later been proved [3]. Next, for $-1< \epsilon < 0$ the spectrum consisted of a few real and the rest as complex-conjugate pairs of discrete eigenvalues. In the former case the energy eigenstates were also the eigenstates [1,2] of PT while the PT-symmetry was exact or unbroken.  Interestingly, $V_{BB}(x,2)=-x^4$ is a real Hermitian barrier (not a well), where the real positive discrete spectrum has been aptly interpreted [4] as the reflectivity zeros in scattering from  flat-top potentials such as $V(x)=-x^{2n+2}$, $n=1,2,3....$

By complexifying Razavy's real potential a quasi-exactly solvable potential was reported [5] displaying
the phenomenon of broken and unbroken PT-symmetry.
However, here $P$ was taken as $x \rightarrow i\pi/2-x$. A real Hermitian and a complex PT-symmetric Scarf II potential were found to have identical [6] spectra. Several other exactly solvable potentials were complexified to produce [7] larger number of exactly solvable
complex PT-symmetric potentials having real discrete
spectrum. Existence of two branches of real discrete
spectrum in complex PT-symmetric Scarf II  was revealed
[8] and interpreted in terms of quasi-parity [9]. Complex
PT-symmetric Scarf II was shown to be an exactly solvable model displaying the spontaneous breaking of PT-symmetry when the  strength of the imaginary part, $|V_2|$, exceeded a critical value of $V_1+1/4$, where $-V_1(V_1>0)$ was the strength of the real part [10].
Such a phase transition of eigenvalues from real to complex conjugate  pairs controlled by a critical parameter has inspired very interesting experiments
in wave propagation [11] and optics where they realized PT-symmetry as equal gain and loss medium. Non-reciprocity of reflection [12] from such complex PT-symmetric mediums has given rise to novel phenomena like spectral singularity [13], coherent perfect absorption (CPA) without [14] and with [15,16] lasing, also see Ref. [17] for exactly solvable models of CPA of both kinds. 

Analyticity of a function in one dimension means continuity and differentiability in a domain. In a very interesting paper [18], the energy eigenspectrum of the potentials like $V(x)=(ix)^a|x|^b$ was studied to  conjecture [18] that, except in rare cases (piecewise  constant potentials [19]), analyticity is an essential feature that is necessary
for [a complex PT-symmetric] Hamiltonian to have the [entire] real spectrum. The words in square brackets were not mentioned earlier which we feel are necessary to be mentioned and emphasized. Following this, it was found that the real Hermitian potentials like $V(x)=x^2$ and $|x|$ having infinite spectrum showed a scarcity [20] of real discrete eigenvalues when perturbed by $V'(x)=ix|x|$ and $ix$, respectively. It turned out [20] that if the total complex PT-symmetric potential is non-analytic, the real discrete eigenvalues may again be found but scarcely. 

In this Letter, we present a new exactly solvable complex PT-symmetric potential which is again non-analytic. We find  both a finite occurrence and a scarcity of 
real discrete eigenvalues, depending upon the values of parameters. Scarcity in the spectrum is shown due to 
the presence of exceptional points (EPs). If $\psi_n(x)$ is the energy-eigenstate, we find that (1): PT$\psi_n(x)=1 \psi_n$  and (2) PT$\psi_{E_n}(x)=\psi_{E^*_n}(x)$ [16,21] when the discrete eigenvalues are real and complex-conjugate pairs, respectively.
The former (latter) is the property of un-broken (broken) PT-symmetry [16,21].   

We consider the one-dimensional potential $V(x)= ig~ \mbox{sgn}(x)~ |1-\exp(2|x|/a)|$ that can be rewritten as
\begin{equation} 
V(x)=\left\lbrace\begin{array}{lcr}
ig[1-e^{-2x/a}], & &  x \le 0\\
ig[e^{2x/a}-1], & & x >  0, \\
\end{array}
\right.
\end{equation}
without a loss of generality, let us assume that $a,g>0$. When $a$ tends to be large, $V(x) \rightarrow 2igx/a$; this is so because $e^z \approx 1+z$ when 
$|z|$ is very small. The  discrete spectrum of $V(x)=i\lambda x$ is known to be null [1].

For $x<0$, the time-independent  Schr{\"o}dinger equation can be written as
\begin{equation}
\frac{d^2\psi}{dx^2}+(p^2+s^2 e^{-2x/a})\psi(x)=0,
\end{equation}
where 
\begin{equation}
p=\frac{\sqrt{2\mu[E-ig]}}{\hbar},\quad s=\frac{1+i} {\sqrt{2}}\frac{\sqrt{2\mu g}}{\hbar}.
\end{equation}
This equation can be transformed into cylindrical Bessel equation [23],
using $y=sae^{-x/a}$. Eq. (2) becomes
\begin{equation}
y^2\frac{d^2\psi(y)}{dy^2}+y\frac{d\psi(y)}{dy}+(y^2+p^2a^2)\psi(y)=0.
\end{equation}
This Bessel equation [23] admits two sets of linearly independent solutions. For our purpose, we seek $H^{(1)}_{ipa}(y)$ and $H^{(2)}_{ipa}(y)$. Using the asymptotic property of cylindrical Hankel functions [23], we find that
\begin{equation}
H^{(1,2)}_{\nu} (y)\sim \sqrt{2 /\pi y}~ e^{[\pm i(y-\nu \pi/2-\pi/4)]}, \quad ~|y| \sim \infty. 
\end{equation}
So the wave functions
\begin{equation}
\psi_1(x) \sim e^{x/2a} e^{isae^{-x/a}} \quad \mbox{and} \quad \psi_2(x) \sim e^{x/2a} e^{-isae^{-x/a}}
\end{equation}
act as asymptotic forms of two linearly independent solutions of
(1) on the left $(x<0)$. Seeing the definition of $s$ in (3) we observe that $\psi_1(x)$ converges  and $\psi_2(x)$ diverges when $x \rightarrow -\infty$.
Thus, for the bound state solution of (1), we write
\begin{equation}
\psi_{<}(x)=\frac{H^{(1)}_{ipa} (sae^{-x/a})}{H^{(1)}_{ipa} (sa)},\quad x<0.
\end{equation}
For $x>0$, by inserting the potential (1) in Schr{\"o}dinger equation
\begin{equation}
\frac{d^2\psi}{dx^2}+(q^2-s^2 e^{2x/a})\psi(x)=0,
\end{equation}
where $q=\frac{\sqrt{2\mu(E+ig)}}{\hbar}$. Then using the transformation  $z=sa e^{x/a}$, we get 
the modified cylindrical Bessel Equation given below:
\begin{equation}
z^2\frac{d^2\psi(z)}{dz^2}+z\frac{d\psi(z)}{dz}-(z^2-q^2a^2)\psi(z)=0. 
\end{equation}
This second order differential equation has two linearly independent solutions $I_{\nu}(z),K_{\nu}(z)$. Most importantly, we choose the latter noting that $K_{\nu}(z) \sim \sqrt{\frac{\pi}{2z}} e^{-z}$
when $z \sim \infty$.
So, we write
\begin{equation}
\psi_{>}(x) = \frac{ K_{iqa}(sae^{x/a})}{K_{iqa}(sa)},\quad x>0.
\end{equation}
The solutions (7) and (10) are self-matched at $x=0$, by matching their derivative at $x=0$ we get the energy eigenvalue formula
\begin{equation}
f(E)={H^{(1)}}_{ipa}' (sa) K_{iqa}(sa)+H^{(1)}_{ipa} (sa) K_{iqa}'(sa)=0,
\end{equation}
for the real discrete spectrum of (1).
For scattering states (positive energy continuum), we seek the solution of (4) when $x<0$ as
\begin{eqnarray}
\psi(x)= A  e^{p \pi a/2} e^{-i\pi/4} H^{(2)}_{ipa}  (sae^{-x/a}) + B   e^{-p \pi a/2} e^{i\pi/4}H^{(1)}_{ipa} (sae^{-x/a})
\end{eqnarray}
\begin{equation}
\psi(x) = C K_{iqa}(sae^{x/a}),\quad x>0.
\end{equation}
By matching the solutions (12) and (13) at $x=0$, we obtain the reflection amplitude ($B/A)$=
\begin{equation}
r(E)=i\left( \frac{ K'_{iqa} (sa)H^{(2)}_{-ipa}(sa)+K_{iqa}(sa) H^{(2)'}_{-ipa}(sa)}{ K'_{iqa} (sa) H^{(1)}_{ipa}(sa)+K_{iqa}(sa) {H^{(1)}_{ipa}}'(sa)} \right).
\end{equation}
The energy function, $f(E)$,  turns out to be the denominator of the reflection amplitude signifying that real energy poles of (14)
are nothing but the real discrete eigenvalues of the exponential potential (1). The results 
(11) and (14) hold for $g>0$. For $g<0$, in these results $p$ and  $q$; the Hankel functions $H^{(1)}$  and  $H^{(2)}$ need to be interchanged. Eventually, eigenvalues of (1) are symmetric, in
this regard $E_n(-g)=E_n(g).$ 

The real discrete spectra of (1) can be obtained by the real zeros of $f(E)=0$, however, finding them by locating the real discrete energies, where the reflection probability $(R(E)=|r(E)|^2)$ becomes infinity, is more interesting and presentable (see Figs. 1). Usually, the real energy, where reflection, $R(E)$, and transmission, $T(E)$, probabilities become infinite is called a spectral singularity [13] for a scattering potential which is such that $V(\pm \infty)=0$. Here, as the potential diverges to $\pm i \infty$, the spectral singularities are the positive energy bound states. A recent study [22]  of the common negative-energy singularities of $R(E)$ and $T(E)$ is shown to yield the real discrete spectrum
of three parametric domains of the complex Scarf II which is neither PT symmetric nor (apparently) pseudo-Hermitian.

In every numerical calculation in the sequel we shall be using $2\mu = 1 = \hbar^2$.
In Fig. 1, we plot $R(E)$ for $ g=1$
and $a=0.5, 1, 2, 5$ in (a,b,c,d), respectively.  
The poles in $R(E)$ at discrete real energies (eigenvalues: roots of Eq. (11)) appear as sharp peaks, just after the peaks the minima indicate coalescing of a pair of real eigenvalues into complex- conjugate pairs. The number of eigenvalues reduces from 9 to just 1 as we increase $a$ from 0.5 to 5 in Fig. 1. In Fig. 1(b), we see five clear spikes at real energies $E=3.27651 (E_0)$,  $8.83705 (E_1)$, $13.7572 (E_2)$, $21.3361 (E_3)$,  $25.6883 (E_4)$ and one wide maximum at $E= 37.5832 \pm 2.6879 i (E_{5,6})$.  We find that the bound eigenstates for real energies are like $\psi(x)=\phi_e(x)+i\phi_o(x)$,
$\phi_{e,o}(x)$ are even and odd parity functions satisfying $\phi_{e,o}(\pm \infty)=0$ [24]. In Fig. 2, we plot real and imaginary parts of $\psi(x)$ for $E=E_0$ and $E_1$. The even (solid line) and odd (dashed line) parities of these parts testify to 
\begin{equation}
\mbox{PT}\psi_n(x)=\psi_n^*(-x)=+1 \psi_n(x), 
\end{equation}
signifying that PT-symmetry is exact (unbroken).  The next $E_5$ and $E_6$ the poles of $R(E)$ or the solution of $f(E)=0$ are the complex-conjugate pairs of energy eigenvalues. The interesting signature of the spontaneous breaking of PT-symmetry results in the loss of definite parities of the real and imaginary parts of the eigenstates $\mbox{PT}(\psi(x))$ or $\psi(x)$ and both differ from each other (see Fig. 3). In  the case of spontaneous breaking of PT-symmetry, the flipping  [16,21] of eigenstates takes place as
\begin{equation}
\mbox{PT}\psi_E(x)=e^{i\alpha} \psi_{E^*}(x), \quad \mbox{PT}\psi_{E^*}=e^{-i\alpha} \psi_{E}(x), ~\alpha \in {\cal R},
\end{equation}
For the present model, see Fig. 3, for the eigenstates $\psi_5(x), \psi_6(x)$ corresponding to complex-conjugate eigenvalues, Figures (a) and (b) are mirror images of each other about $y-$axis and figures (c) and (d) are minus times the mirror image of each other about $y-$axis. Thus, in the case of the exponential potential (1), we have $\alpha=0$ in (16).

For $a=1$, we find three consecutive exceptional points
(values of $g$) as 0.74, 1.58, 5.35. The least number of real discrete eigenvalues is 7 when $g<0.74$. Exact numbers of real discrete eigenvalues for $0.74 < g <1.58$,  $1.58 <g <5.35$ and $g>5.34$ are 5, 3 and 1,  respectively (see Fig. 4). Additionally, for $g=0.1$ and $g=0.2$, we get 12 and 9 eigenvalues, respectively. 
Similarly, for $a=0.5$, we find four consecutive exceptional points as $g= 1.62, 2.96, 6.29, 21.38$ (see Fig. 5). When $g<1.62$ there are at least 9 eigenvalues. Then for $1.62 <g< 2.96$, $2.96 < g < 6.29$, $6.28 <g <31.38$, in Fig. 5, one can see that there are 7, 5, 3 and 1 eigenvalues, respectively.
When the parameters $a$ and $g$ are very small there may be an abundance of real eigenvalues but the entire discrete spectrum cannot be real. We conjecture that in the proposed non-analytic complex PT-symmetric exponential potential the entire discrete spectrum cannot be real. For higher values of $a$ and $g$, there will be scarcity of real discrete eigenvalues; nevertheless
there will be at least one real discrete eigenvalue but for the limiting case of
$a\rightarrow \infty$ (when $V(x)$ (1)) it passes over to $i\lambda x$. 

The behaviour $E_n(g)$ for $g \in (g_2,g_3)$ in Fig. 4
and for $g \in (g_3,g_4)$ in Fig. 5 where there are only three real discrete eigenvalues, is akin to that of the quasi-exactly solvable complex PT-symmetric potential for $M=3$ [5]. There the 
ground state is un-paired with $E_0=5-z^2$ but $E_1^{\pm}= 7-z^2\pm 2\sqrt{1-4z^2},0<z<1/2$.

Strangely, despite having analytic eigenstate given by Eqs. (7) and (10), it is indeed challenging to prove  elegant properties (15) and (16) of  energy eigenstates in PT-symmetry. Just to bring out the level  of difficulty, we at best can express [23]
\begin{eqnarray}
H^{(1)}_\nu(z)&=&\frac{2}{\sqrt{\pi}} e ^{[-i(\pi(\nu+1/2)-z]}(2x)^\nu U(\nu+1/2, 2\nu+1; -2iz),\nonumber \\
K_\nu (z)&=&\sqrt{2} e^{-z} (2z)^\nu U(\nu+1/2,2\nu+1;2z). 
\end{eqnarray}
Here $U(a,b;z)$ is second Kummer's function which is solution of confluent hyper-geometric equation with limited properties [23].
The argument of these functions in our case is like $z=(s_1+is_2)e^{x/a}$ under PT-symmetry it will change to 
$z=(s_1-is_2)e^{-x/a}.$ Such a transformation of $U$ does not seem to be available.

Recently, it has been shown that $L^2-$square integrable eigenstates of a complex potential satisfying Dirichlet boundary condition $\psi(\pm \infty)=0$,  regardless of real or non-real discrete eigenvalues, are orthogonal as [22]
\begin{equation}
\int_{-\infty}^{\infty} \psi_m(x) \psi_n(x) dx =0.
\end{equation}
We find that the same holds true here as well. For instance, 6 eigenstates corresponding to the case when
$g=1$ and $a=1$ (Figs. 1(b), 2, 3) are mutually orthogonal as (18). In case we use PT-scalar product as $ <\psi_m(-x)|\psi_n(x)>$ [2,9,10], then due to the property (16), $\psi_5(x)$ and $\psi_6(x)$ corresponding to complex-conjugate eigenvalues are self- orthogonal and they have PT-norm as zero.

We conclude by asserting that we have presented a new solvable complex PT-symmetric non-analytic exponential potential which has both a finite occurrence and a scarcity (existence of exceptional points) of real discrete spectrum depending on the values of its parameters. Based on our calculations, we conjecture that this potential cannot have entire  discrete spectrum as real. In other words, there is no parametric separation of unbroken and broken PT-symmetry. There exists exceptional points (EPs) (several critical values of the parameter $g$ for fixed values of $a$) below which there are only a finite number of real discrete eigenvalues. However, the states having real eigenvalue are also {\it degenerate} eigenstates of PT all having eigenvalue of PT as 1. It may be recalled that for PT-symmetric Scarf II, these are $(-1)^n$.  Next, in the case of complex-conjugate pairs of eigenvalues, the action of PT flips the eigenstate $\psi_E(x)$ to $\psi_{E^*}(x)$. This potential has at least one real eigenvalue for finite values of $a$.  An interesting anti-climax of the new solvable model is that the commonly known properties of Hankel and modified Bessel functions fall short to prove the elegant properties  (15,16) of eigenstates under the action of PT; we have confirmed them numerically. This physical requirement may inspire one to look for some new or rarely known properties of these aforementioned higher order functions.\\
\section*{\Large{Acknowledgements}}
We thank  Dr. V.M. Datar for his support and interest 
in this work.

\begin{figure}[H]
\centering
\includegraphics[width=7 cm,height=5. cm]{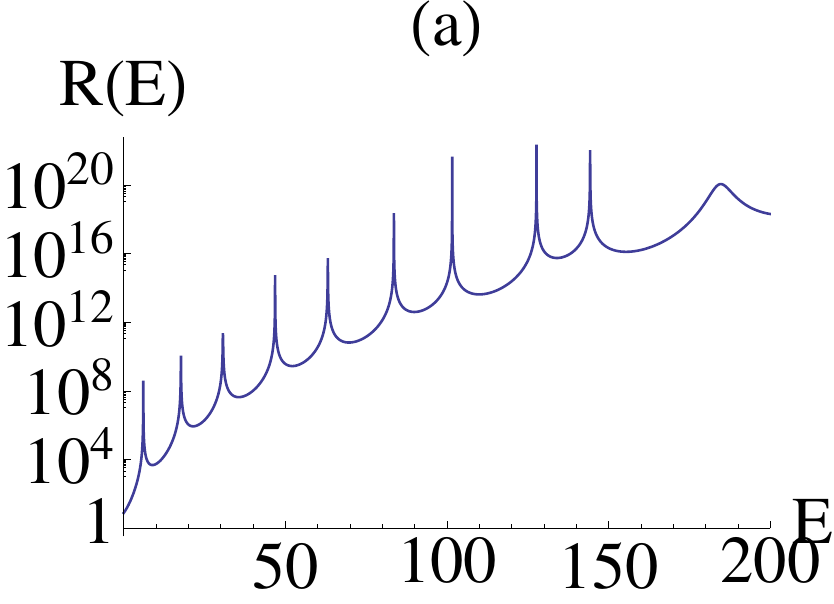}
\hskip .5cm
\includegraphics[width=7 cm,height=5. cm]{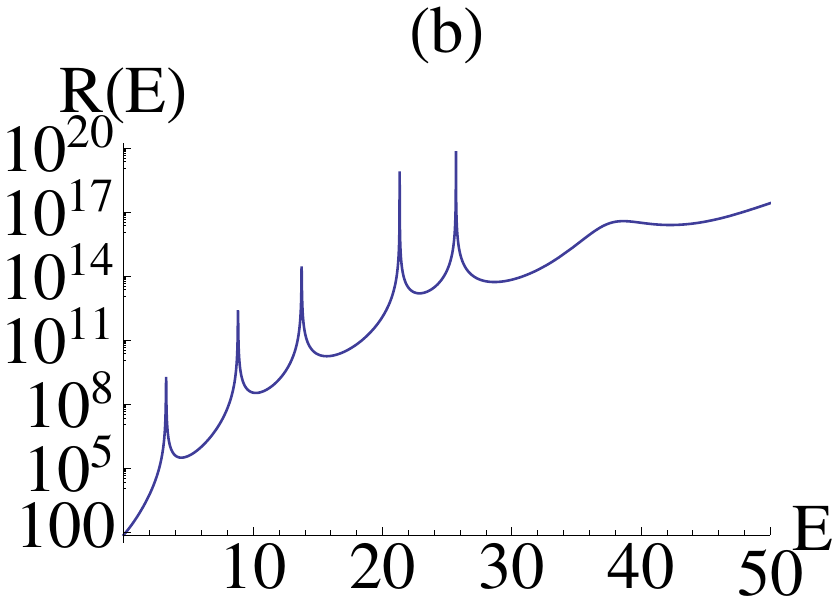}
\vskip .5cm
\includegraphics[width=7 cm,height=5. cm]{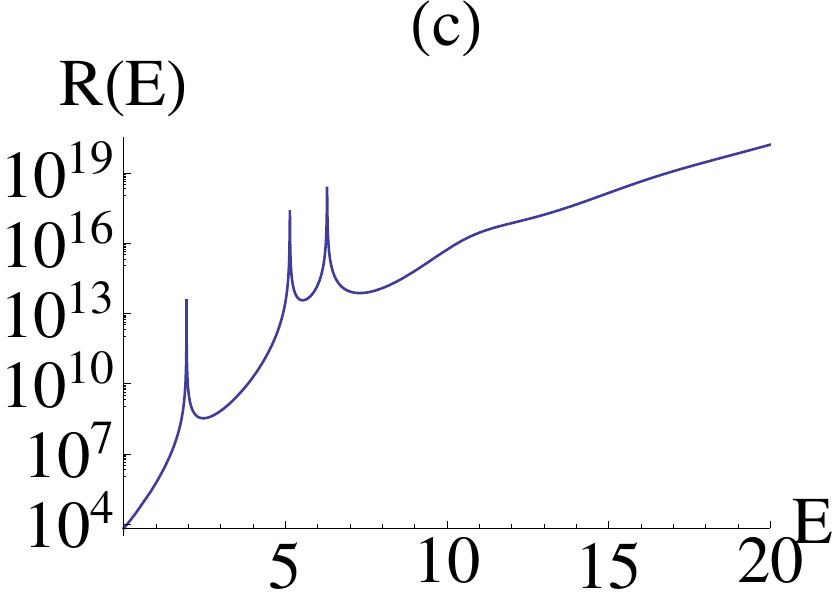}
\hskip .5 cm
\includegraphics[width=7 cm,height=5. cm]{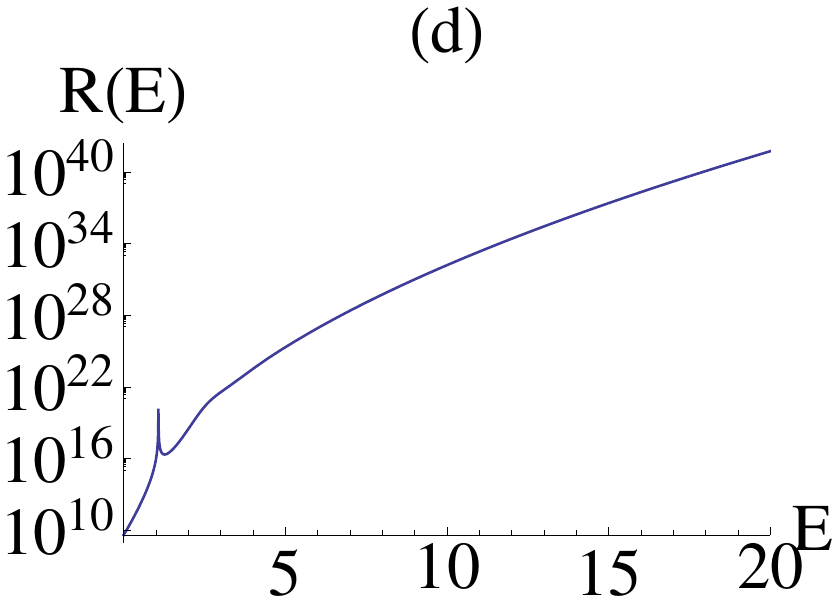}
\caption{Spikes in reflectivity, $R(E)$ (14) indicating real discrete eigenvalues (solutions of Eq.(11))  of the newly proposed potential (1). Here $g=1$ and $a$ is being varied: (a) 0.5, (b) 1.0, (c) 2.0, (d) 5.0.
In (a,b,c),  see a hump after discrete energy poles  indicating the next two eigenvalues have merged 
to become complex-conjugate pairs. One can actually choose a value of $g$ for which the last two discrete energies will be very close by as real doublets.  Notice that as $a$ increases, the number of real eigenvalues reduces from 9 to just 1. In fact, for very large values of $a$ there will be no real eigenvalue. This is so because the potential (1) passes over to $V(x)=i\lambda x$ which is known to have  discrete spectrum as null.}
\end{figure}
\begin{figure}[H]
\centering
\includegraphics[width=7 cm,height=5. cm]{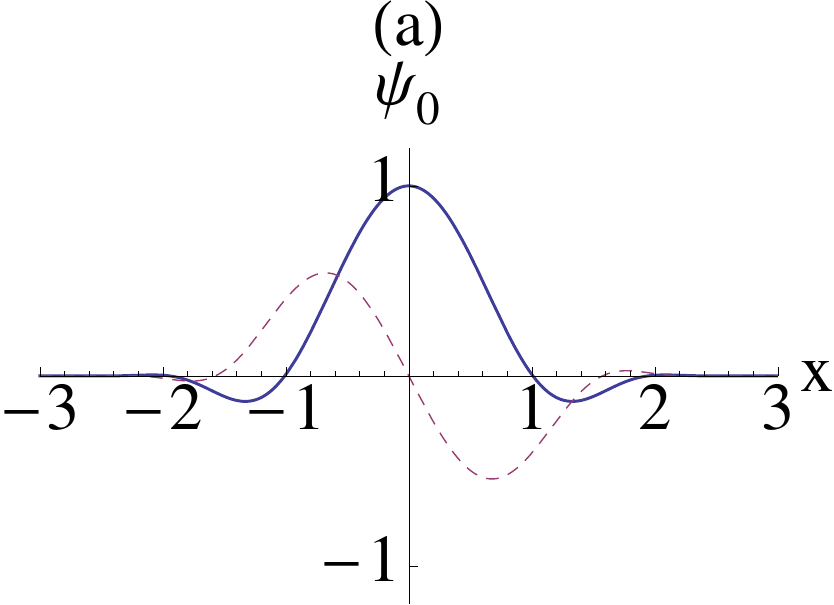}
\hskip .5cm
\includegraphics[width=7 cm,height=5. cm]{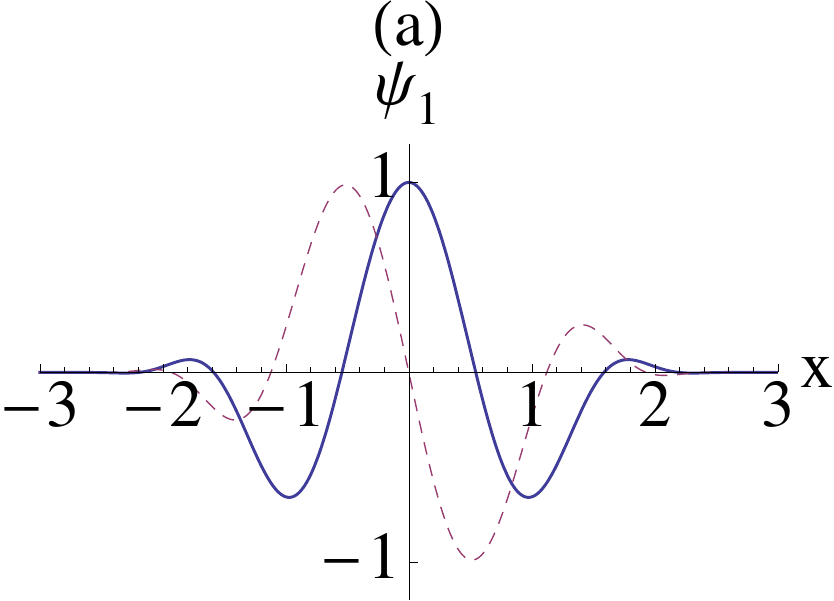}
\caption{The typical complex eigenstates corresponding to real eigenvalues (a) $E_0=3.27651$ and  (b) $E_1=8.83705$. The solid (dashed) lines show the real (imaginary) parts. Here $a=1$ and $g=1$ as in Fig. 1(b). In general, we find that  these eigenstates are square integrable of the type: $\psi_n(x)=\phi_{e,n}(x)+i\phi_{o,n}(x), ``e(o)"$ denote even(odd).}
\end{figure}
\begin{figure}[H]
\centering
\includegraphics[width=7 cm,height=3. cm]{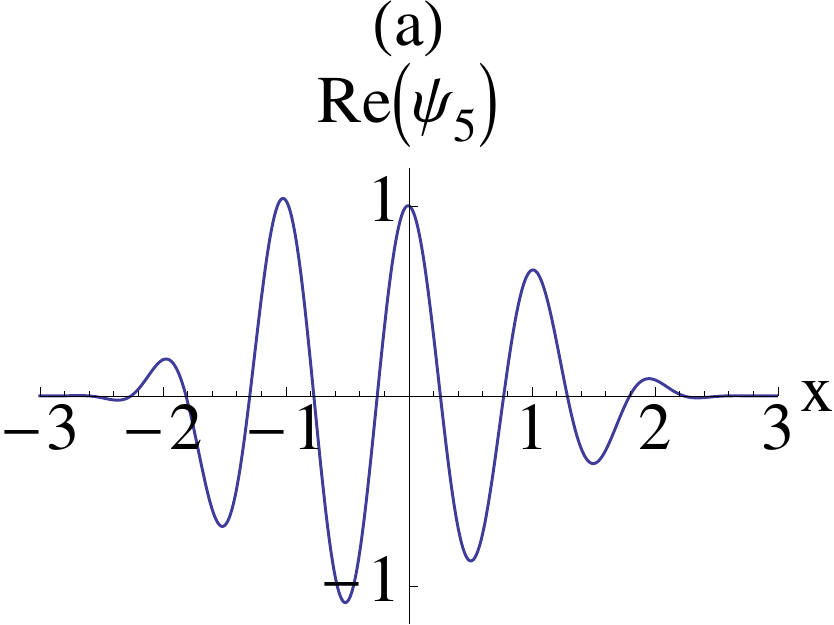}
\hskip .5cm
\includegraphics[width=7 cm,height=3. cm]{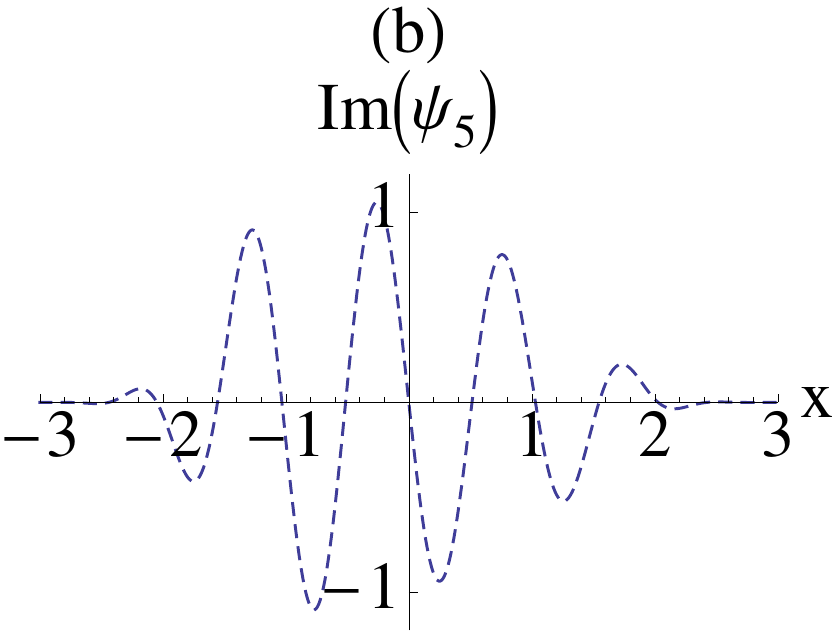}
\vskip .5cm
\includegraphics[width=7 cm,height=3. cm]{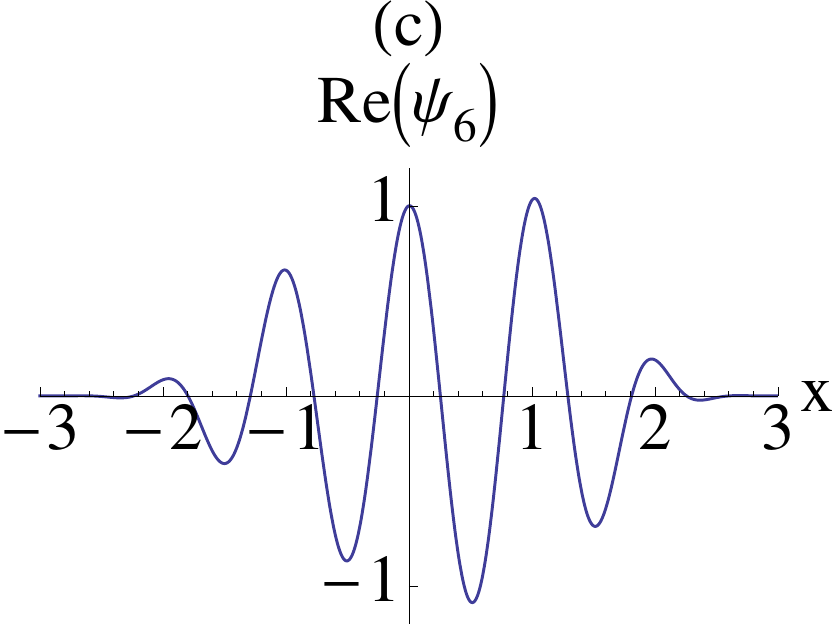}
\hskip .5 cm
\includegraphics[width=7 cm,height=3. cm]{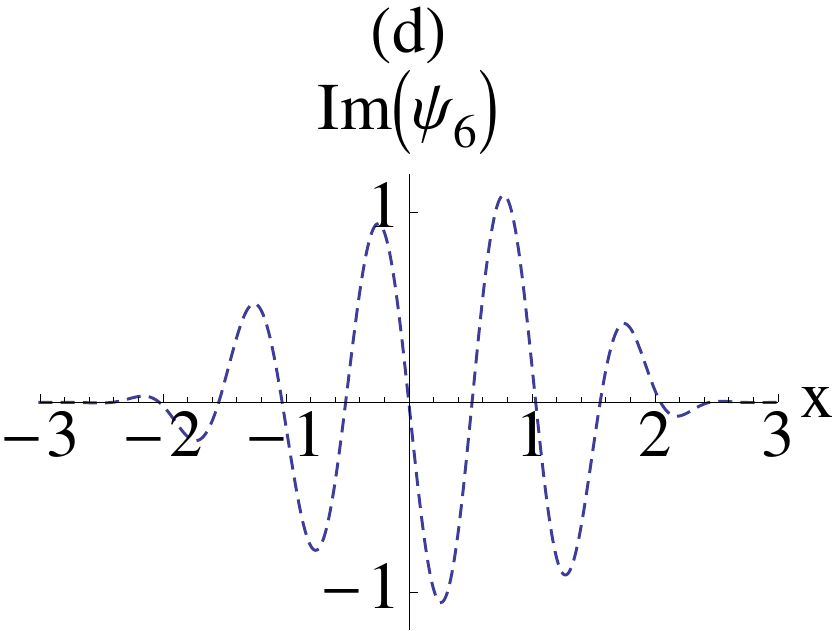}
\caption{Typical eigenstates $\psi_5(x)$ and $\psi_6(x)$ corresponding to discrete complex-conjugate energy eigenvalues, $37.5832 \pm 25.6883 i$, respectively. Potential parameters are same as that of Figs. 1(b) and 2. Notice that these eigenvalues appear in Fig. 1(b) as a hump after 4 real discrete energy peaks (poles). One can closely look into these figures to make out the acclaimed result, i.e., $\mbox{PT}(\psi_{5}(x))= \psi_6(x)$.  More directly, see that the figures (a) and (c) are reflection
of each other about $y-$axis, but (b) and (d) are minus times the mirror reflection of the each other. Numerically,
$\psi_{5}(1)=0.661638 + 0.121078 i, \psi_{5}(-1)=1.02862 - 0.201041 i$ and $\psi_{6}(1)=1.02862 + 0.201041 i, \psi_6(-1)=0.661638 - 0.121078 i$.}
\end{figure}
\begin{figure}[H]
\centering
\includegraphics[width=18 cm,height=8. cm]{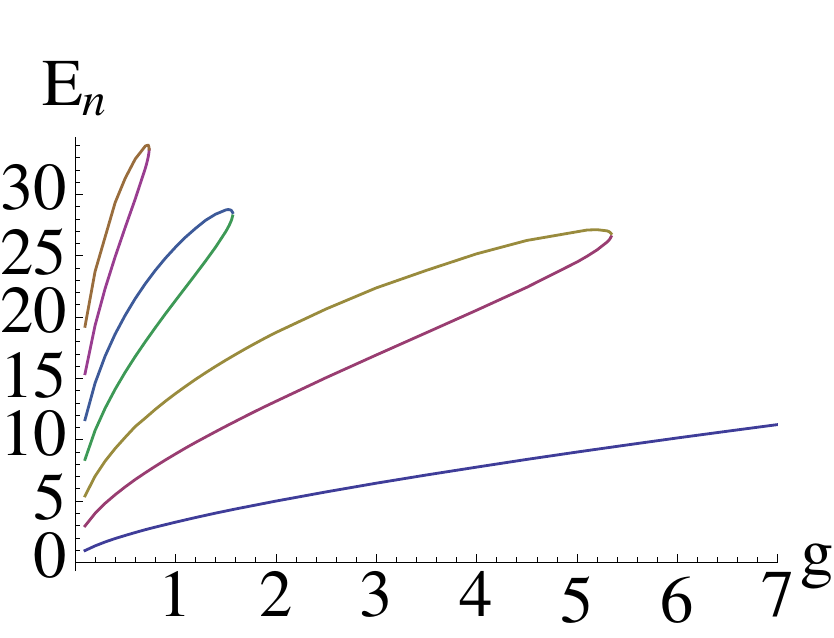}
\caption{Real discrete energy eigenvalues of the exponential potential (1) as a function of $g$ for $a=1.0$: there are three EPs: $g_1=0.74, g_2=1.58, g_3=5.35$.}
\end{figure}
\begin{figure}[H]
\centering
\includegraphics[width=18 cm,height=8. cm]{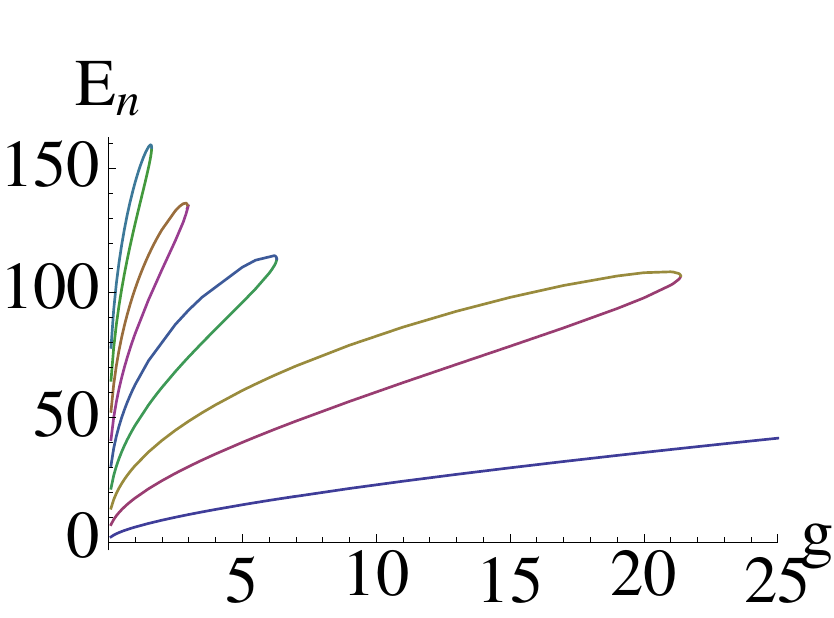}
\caption{Real discrete energy eigenvalues of the exponential potential (1) as a function of $g$ for $a=0.5$: there are three EPs: $g_1=1.62, g_2=2.96, g_3=6.29, g_4=21.38$.}
\end{figure}
\end{document}